

\input harvmac

%
\def\np#1#2#3{Nucl. Phys. B{#1} (#2) #3}
\def\pl#1#2#3{Phys. Lett. {#1}B (#2) #3}

\def\prep#1#2#3{Phys. Rep. {#1} (#2) #3}

\def\Lam#1{\Lambda_{#1}}
\def\pf{{\rm Pf ~}}
\font\zfont = cmss10 
\font\litfont = cmr6

\def\bigone{\hbox{1\kern -.23em {\rm l}}}
\def\ZZ{\hbox{\zfont Z\kern-.4emZ}}
\def\half{{\litfont {1 \over 2}}}
\def\mx#1{m_{\hbox{\fvfont #1}}}

\def\gG{{\cal G}}
\def\lamlam#1{\langle S_{#1}\rangle}
\def\lfm#1{\medskip\noindent\item{#1}}

\Title{hep-th/9403198, RU-94-26}
{\vbox{\centerline{Exact Superpotentials in Four Dimensions}}}
\bigskip
\centerline{K. Intriligator, R.G. Leigh and N. Seiberg}
\smallskip
\centerline{\it Department of Physics and Astronomy}
\centerline{\it Rutgers University, Piscataway, NJ 08855-0849}
\bigskip
\baselineskip 18pt
\noindent
Supersymmetric gauge theories in four dimensions can display interesting
non-perturbative phenomena.  Although the superpotential dynamically
generated by these phenomena can be highly nontrivial, it can often be
exactly determined.  We discuss some general techniques for analyzing
the Wilsonian superpotential and demonstrate them with simple but
non-trivial examples.

\Date{3/94}

\newsec{Introduction}

There are three motivations to study supersymmetric field theories.
First, theories with dynamical supersymmetry breaking can be used to solve
the hierarchy problem.
Second, they are relevant to topological field theories.
Finally, they are tractable and can thus be used as testing
grounds for various ideas about the dynamics of four dimensional quantum
field theories.

In four-dimensional quantum field theory, exact results, aside from
those which follow directly from symmetries, are very hard to come by.
Supersymmetric theories, however, are different.  The combination of the
holomorphy of the Wilsonian superpotential, $W_{eff}$, with the
symmetries and selection rules provides powerful constraints.  These
constraints should be viewed as ``kinematics.''  When combined with
approximate dynamical information about the asymptotic behavior of the
superpotential, we can sometimes determine it exactly
\ref\nonren{N. Seiberg, \pl{318}{1993}{469}.}.

In this paper we continue this line of reasoning and apply it to more
complicated systems.  Unlike the models analyzed in ref. \nonren, where
the $W_{eff}$ turned out to be rather simple functions, here we find
highly non-trivial effective superpotentials.  These reflect interesting
new non-perturbative effects.

\lref\dds{A.C. Davis, M. Dine and N. Seiberg, \pl{125}{1983}{487}.}
\lref\ads{I. Affleck, M. Dine, and N. Seiberg, \np{241}{1984}{493};
\np{256}{1985}{557}.}
\lref\cerne{G. Veneziano and S. Yankielowicz, \pl{113}{1982}{321};
T.R. Taylor, G. Veneziano, and S. Yankielowicz, \np{218}{1983}{493}.}
\lref\cern{D. Amati, K. Konishi, Y. Meurice, G.C. Rossi and G. Veneziano,
\prep{162}{1988}{169} and references therein.}

We will always be interested in the Wilsonian effective action.  If
supersymmetry is broken we limit ourselves to scales above the breaking
scale,
where supersymmetry is linearly realized.  We will integrate out the
massive modes and focus on the dynamics of the light fields.  In
this respect we follow the point of view of refs. \refs{\dds,\ads}. An
alternate approach \refs{\cerne,\cern} uses an effective Lagrangian
which also depends on some of the massive fields.
We discuss generally how to integrate these massive fields into the low
energy theory.

In section 2 we summarize our techniques.  The low energy superpotential
is constrained by the symmetries and holomorphy as in ref. \nonren. The
dynamical analysis can proceed in two different ways: we can analyze the
asymptotic behavior of the superpotential in various limits, control its
singularities, and thus completely determine it.  Alternatively, we can
derive differential equations that the superpotential satisfies as a
function of various coupling constants and thus solve for it.

In section 3 we give a brief review of the dynamics of supersymmetric
QCD.  Sections 4 and 5 are devoted to examples demonstrating our
techniques.

In section 4 we study an $SU(2)_1\times SU(2)_2$ gauge theory with
matter fields in the representations $Q=({\bf 2},{\bf 2})$ and
$L_i=({\bf 1},{\bf 2})$ for $i=1,\dots ,2n$.  In terms of the gauge
singlet composites $X=Q^2$ and $V_{ij}=L_iL_j$, we find the
superpotentials
\eqn\suiiws{\eqalign{W_{n=0}&={\left(\Lambda _1^{5/2}\pm \Lambda
_2^{5/2}\right)^2\over X}\cr
W_{n=1}&={\Lambda _1^5\; V_{12}\over XV_{12}-\Lambda _2^4}\cr
W_{n=2}&=-{X\;\pf V\over \Lambda _2^3}\pm 2\;
\sqrt{{\Lambda_1^5\;\pf V\over\Lambda_2^3}},\cr}}
where $\Lambda _1$ and $\Lambda _2$ are the scales of the $SU(2)_1$ and
$SU(2)_2$ gauge theories, respectively.  Note that for $n=2$
the fields $V$ are classically constrained by $\pf V=0$.  However, quantum
mechanically, $\pf V$ is a massive field whose expectation value
satisfies $\langle\pf V\rangle = {\Lambda_1^5 \Lambda_2^3 \over X^2}$.
The low energy effective Lagrangian, after $\pf V$ is integrated out, is
\eqn\wtild{\tilde W_{n=2} = {\Lambda_1^5 \over X}  .}

The $\pm$ signs in eqs. \suiiws\ label distinct low energy groundstates,
differing in the expectation
value of a massive field which is not included in the low energy
effective action. The $\pm$ sign in the superpotential $W_{n=2}$
in \suiiws\ corresponds to two different branches of the square root; they
are related by a discrete symmetry of the theory and
therefore describe equivalent physics. On the other hand, the $\pm$
sign in the superpotential $W_{n=0}$ in
\suiiws\ labels two inequivalent (unrelated by a symmetry) low energy
groundstates. The low energy theory includes then both continuous
fields and discrete labels.  A similar phenomenon was observed in ref.
\ref\irratio{T. Banks, M. Dine and N. Seiberg, \pl{273}{1991}{105}.}.

In section 5 we consider an $SO(5)\times SU(2)$ gauge theory with a
matter field in the representation $F=({\bf 4},{\bf 2})$ with or
without two fields $L_{1,2}=({\bf 1},{\bf 2})$.  In terms of the gauge
singlet fields $X=F^2$ and $Y=L_1L_2$, we find the superpotentials to be
\eqn\oivw{\eqalign{
W_{n=0}&= {2\Lambda _5^4\over \sqrt{X\pm 2\Lambda _2^2}}\cr
W_{n=1}={\Lambda _5^4\over \sqrt{X}}\; g\left(v={\Lambda _5^4
\Lambda _2^3\over X^{5/2}Y}\right)\ \hbox{with}\  &
g=\half h(5-h^2)\ \hbox{and}\ v=\half (h^{-3}-h^{-5}).}}
For $W_{n=1}$ we were unable to
find a closed form expression from the parametric solution in \oivw .
The sign choice in $W_{n=0}$ in \oivw\ is, again, a discrete
label in the low energy theory.  The groundstates differing by this sign
choice are here related by a symmetry and are thus physically equivalent.

As is clear from eqs. \suiiws\ and \oivw , the superpotentials are quite
complicated.  They are generated by a variety of dynamical mechanisms.
For example, the large field behavior of $W_{n=1}$ in \suiiws\ arises
{}from an infinite sum over instantons whereas the large field behavior of
$W_{n=0}$ in \oivw\ arises from an interplay between gaugino condensation in
the two groups and an infinite number of instantons.
The dynamics leading to the superpotential generally depends upon
the  region of  field space considered; the holomorphic superpotential
smoothly interpolates between them.

We conclude in section 6 with an outlook and various speculations.

\newsec{Techniques}

Our general framework is a supersymmetric field theory based on a gauge
group $\gG$ and matter superfields $\phi^i$ transforming in representations
$R_i$ of $\gG$.  The tree level superpotential is
\eqn\treew{W_{tree}= \sum_r g_r X^r(\phi^i), }
where $X^r$ are gauge invariant polynomials in the fundamental fields.
Apart from the tree level couplings $g_r$, we also have gauge couplings:
every simple factor $\gG_s$ in $\gG = \prod _s \gG_s $ is characterized
by a scale $\Lambda_s$.

Our analysis proceeds in several steps:

\smallskip
\noindent
I. ~~~ We first set the tree level superpotential to zero, {\it i.e.}\
$g_r=0$. At the classical level there are then ``flat directions'' in
field space where all the gauge $D$-terms vanish.  The expectation
values of the scalar components of $\phi^i$ in these classical ground states
spontaneously break the gauge symmetry.  We refer to this space of
classical ground states as ``the classical moduli space.''  Instead of
using the fundamental fields $\phi ^i$ as coordinates on this space, we
can use gauge invariant combinations $X^r$.  The $X^r$ are the
light superfields in the leading approximation; the classical low energy
superpotential for them vanishes. It is sometimes the case that the
fields $X^r$ are constrained classically
\ref\nati{N. Seiberg, Rutgers preprint RU-94-18, hep-th/9402044.}.
In this situation, we can represent the constraint with a Lagrange
multiplier in the effective superpotential.

\smallskip
\noindent
II. ~~~ Next we turn on the coupling constants $g_r$ and
$\Lambda_s$; {\it i.e.}\ consider the full quantum theory.  If it is
clear that some of the fields $X^r$ are massive we can either keep them
in our description or look for an effective Lagrangian after they have
been integrated out. The full quantum superpotential $W_{eff}$ is
constrained by two kinematic constraints \nonren:

\nref\russ{M.A. Shifman and A.I Vainshtein, \np{277}{1986}{456};
\np{359}{1991}{571}.}
\lfm{1.} Holomorphy:  $W_{eff}$ is a holomorphic function of the fields
$X^r$ and the coupling constants $g_r$ and $\Lambda_s$.  Holomorphy in
the coupling constants follows from thinking about them as background
fields.  A related discussion of holomorphy in coupling constants of
various expectation values may be found in refs. \refs{\cern , \russ}.
\lfm{2.} Symmetries: $W_{eff}$ is invariant under all the symmetries in
the problem.  If a symmetry is explicitly broken by the coupling
constants we can assign transformation laws to these constants such that
$W_{eff}$ is invariant under the combined transformation on the fields
$X^r$ and the coupling constants.  Anomalous symmetries should be viewed
as explicitly broken.  However, by assigning appropriate transformation
laws to the scales $\Lambda_s$ of the gauge groups, they also lead to
selection rules.

\smallskip
\noindent
III. ~~~ The dynamics enters through the analysis of $W_{eff}$ at
various asymptotic values of its arguments.  In ref. \nonren\ the weak
coupling limit of small $g_r$, small $\Lambda_s$, and large fields $X^r$
was powerful enough to completely determine $W_{eff}$.  In our new
examples these constraints do not uniquely determine $W_{eff}$ and
therefore we also need to study other limits.  Among these limits will
be strong coupling and small fields $X^r$.  The key fact is that
$W_{eff}$, by holomorphy, is completely determined by its behavior at
various asymptotics and by its singularities.

A special limit that is often useful is when one of the matter fields is
very heavy.  Its mass $m$ is one of the coupling constants $g_r$.
When it is large the massive field can be integrated out.  We can do
this either in the microscopic gauge theory or in the effective low
energy theory.  The first of these yields a new microscopic gauge theory
with fewer matter fields and whose coupling constants, $g_r$ and
$\Lambda_s$, depend on $m$.  The low energy effective superpotential of
this theory should be the same as the one obtained by integrating out
the appropriate fields in the effective Lagrangian of the original
theory.

It is often the case that the two kinematic conditions and the
dynamics at small $g_r$ constrain the effective superpotential to be of
the form
\eqn\weefsi{W_{eff}= W_{eff}(g_r=0) + \sum_r g_r X^r =
W_{dyn}(\Lambda_s, X^r) + W_{tree},}
{\it i.e.}\ it is linear in the $g_r$'s ($W_{dyn}$ includes the Lagrange
multiplier terms for the various constraints that the composite fields
$X^r$ should satisfy).  This is the case in all of our examples.  When
this is not the case, we conjecture that it is always possible and
natural to re-define the fields $X_r$ as a function of the $g_r$ to bring
the superpotential to the form \weefsi\ (for a related discussion see
ref.
\ref\kaplu{V. Kaplunovsky and J. Louis, UTTG-94-1,LMU-TPW-94-1.}).
Now let us integrate out some field, say $X^0$.  The resulting
superpotential $\tilde W_{eff}$ is obtained by solving
\eqn\intxo{{\partial W_{eff}\over \partial X^0}(\langle X^0\rangle)  =0}
for $\langle X^0\rangle$ as a function of all the other fields $X^r$ ($r
\not= 0$) and all the coupling constants $g_r$, and substituting back
into $W_{eff}$.  Clearly $\tilde W_{eff}$ is not linear in $g_0$.  To
see that it is linear in all the other $g_r$'s, note that
\eqn\linresr{{\partial \tilde W_{eff} \over \partial g_r} = {\partial
W_{eff} \over  \partial g_r}(\langle X^0\rangle) +
{\partial \langle X^0\rangle \over \partial
g_r} {\partial W_{eff} \over \partial X^0}(\langle X^0\rangle)
= X^r  \quad {\rm for~} r\not=0  .}
This suggests the definition of $\tilde W_{dyn}$
\eqn\weefsi{\tilde W_{eff}= \tilde W_{dyn} + \sum_{r\not= 0} g_r X^r}
which depends on the light fields $X^r$  ($r \not= 0$), the scales
$\Lambda_s $ and $g_0$. An equation similar to \linresr\ for $r=0$ is
\eqn\linres{{\partial \tilde W_{eff} \over \partial g_0} ={\partial
\tilde W_{dyn} \over \partial g_0} = \langle X^0\rangle  .}

A slight generalization of the previous discussion involves the gauge
coupling constants.  Unlike the $g_r$, our effective Lagrangians do not
involve any field which couples linearly to the gauge coupling
constants.  The reason for this is that the corresponding fields
$S_s=-(W_\alpha^2)_s$ (with this sign the lowest component of $S_s$ is
$+(\lambda \lambda)_s$) are always massive and thus do not have to be
included in a low energy Wilsonian effective action.  However, by
repeating the previous discussion with $g_0$ replaced by $\ln\Lambda_s
^{n_s}$, where $n_s$ is determined by the one loop beta function ({\it
e.g.}\ for $SU(N_c)$ gauge theory with $N_f$ quark flavors in the
fundamental and antifundamental representations, $n=3N_c-N_f$), we learn
that
\eqn\linresg{ {\partial W_{eff} \over \partial \ln
\Lambda_s^{n_s}} ={\partial W_{dyn} \over \partial \ln
\Lambda_s^{n_s}}  =\langle S_s\rangle .}
In deriving \linresg\ we are assuming that the effective superpotential with
the $S_s$ included is linear in $\ln \Lambda _s^{n_s}$, as with the other
couplings $g_r$ in \weefsi .  This is the case in all our examples and,
as with the other $g_r$, we conjecture that it is always true.

To summarize, we conjecture that at every scale the superpotential has
the form
\eqn\weffsu{W_{eff}= W_{dyn} + \sum_r g_r X^r,}
where $W_{dyn}$ depends on the fields $X^r$ and on the coupling constants
of the fields $X^0$ and $S_s$ which have been integrated out such that
\eqn\wdyndif{\eqalign{
& {\partial W_{dyn} \over \partial g_r} = 0 \quad {\rm for~} r\not= 0
\cr
& {\partial W_{dyn} \over \partial g_0} = \langle X^0\rangle \cr
&{\partial W_{dyn} \over \partial \ln \Lambda_s^{n_s}} =
\langle S_s\rangle . \cr}}
These equations can be used in two different ways:

\lfm{1.}  If we know the expectation values $\langle X^0\rangle$
or $\langle S_s\rangle$ as a function of the other fields and coupling
constants, we can use eq. \wdyndif\ to solve for $W_{dyn}$.  This leads
to differential equations for the superpotential.

\lfm{2.}   If we know the $g_0$ dependence of $W_{dyn}$ at some scale,
we can find the expectation value of the massive field $\langle
X^0\rangle$ and using this information we can find the superpotential
before it has been integrated out (we will refer to this procedure as
``integrating in'').

As explained in point 2 above, we can use the ``integrating in''
procedure to construct an effective Lagrangian similar to that of ref.
\cerne\ involving the massive fields $S_s$.   However, since the fields
$S_s$ are always massive and our effective actions are Wilsonian, the
meaning of such an effective action for the $S_s$ is not clear to us.  Its
only virtue is that it allows one to determine the
$\langle S_s\rangle$ by their equations of motion.

The third equation in \wdyndif\ allows us to derive the Konishi
anomaly \cern\
\eqn\konishi{\left<\phi^i { \partial W_{tree} \over \partial \phi^i}
\right>-\sum_s \mu_i^s \lamlam{s}=0 \quad {\rm for~every~}i,}
where $\mu_i^s$ is the index of the representation of the
field $\phi^i$ under the $\gG_s$ gauge group.  To do this, consider the
$U(1)_i$ transformation $\phi^i \rightarrow e^{i \alpha} \phi^i$ under
which the composite field $X^r$ has charge $q^i_r$.  This symmetry is
broken both by the coupling constants $g_r$ and by the anomaly.
However, if we also assign charge $-q^i_r$ to $g_r$ and charge $\mu_i^s$
to $\Lambda_s^{n_s}$, the invariance of the superpotential states that
\eqn\winv{\sum_r q^i_r X^r {\partial W_{eff} \over \partial X^r} -
\sum_r q^i_r g_r {\partial W_{eff} \over \partial g_r} + \sum_s \mu_i^s
 \Lambda_s^{n_s} {\partial W_{eff} \over \partial \Lambda_s^{n_s}}=0 .}
Using eqs. \weffsu\ and \wdyndif ,
\eqn\wdynin{\sum_r q^i_r X^r { \partial W_{dyn} \over \partial X^r} +
\sum_s \mu_i^s \langle S_s\rangle =0 .}
Imposing the equations of motion  ${\partial W_{eff} \over
\partial X^r} (\langle X^r\rangle) =0$, this leads to
\eqn\kone{\sum_r q^i_r g_r\langle X^r\rangle=\sum_s \mu_i^s \langle
S_s\rangle }
which is equivalent to \konishi. Note that \wdynin\ applies
more generally to off-shell $X_r$.

\newsec{Review of a simple example: supersymmetric QCD}

In this section we illustrate some of our basic ideas and conventions in
the context of a well studied example: supersymmetric $SU(N_c)$ gauge
theory with $N_f$ flavors of matter superfields $Q_{cf}$ and $\tilde
Q^{cf}$ in the representations {\bf N}$_c$ and $\bar {\rm\bf N}_c$,
respectively, of $SU(N_c)$.

\subsec{Kinematics: symmetries and holomorphy}

The exact Wilsonian effective superpotential for supersymmetric QCD is
completely determined by the symmetries along with holomorphy.
The superpotential can only depend on the the combination of fields
$\Delta \equiv \det _{ff'}(Q_{cf}\tilde Q^{cf'})$, the unique gauge
singlet which is also a singlet under the $SU(N_f)_L\times SU(N_f)_R$
global flavor symmetry.  For each flavor $f$ there are symmetries
$U(1)_{Q_f}$ and $U(1)_{\tilde Q_{^f}}$ which count the superfield
$Q_{cf}$ or $\tilde Q^{cf}$, respectively, with charge one and all other
fields with charge zero.  In addition there is an R-symmetry $U(1)_R$
under which squarks have charge zero, the quark components of the chiral
superfields have charge $-1$, and the gauginos have charge $+1$.  The
charge conjugate fields, which make up the anti-chiral superfields, of
course have the opposite charges under all these symmetries.

Quantum mechanically, one linear combination of the above $U(1)$
currents is anomalous.  Rather than finding linear combinations for
which the anomaly cancels, it is possible to use the anomaly to find
selection rules.  Following the spirit of \nati , we think of $Y={8\pi
^2\over g^2}+i\theta$, which is the coupling for $S$, as a background
chiral field.  It is seen that the anomaly in each of the $U(1)$
transformations can be canceled by combining them with a transformation
of $\Lambda _{N_c,N_f}^{3N_c-N_f}=\mu ^{3N_c-N_f} e^{-Y(\mu )}$ (the
exponent is given exactly in our Wilsonian treatment by the one loop
beta function \russ ).  The charge to be assigned to the scale in order
to cancel the anomaly is related to the charge assignments of the quarks
$\psi _{cf}$ and $\tilde \psi ^{cf}$ and the gauginos $\lambda$ by
\eqn\cnclanom{ q(\Lambda _{N_c,N_f}^{3N_c-N_f})=\sum _f(q(\psi
_{cf})+q(\tilde \psi^{cf}))+2N_c\; q(\lambda).}

The exact superpotential must have charge zero under the $2N_f$
symmetries $U(1)_{Q_f}$ and $U(1)_{\tilde Q_f}$ and have charge two (for
the lowest component) under the R-symmetry $U(1)_R$.  $\Delta$ has
charge one under each of the $2N_f$ $U(1)$ symmetries and it has zero
R-charge.  Using \cnclanom , $\Lambda _{N_c,N_f}^{3N_c-N_f}$ also has
charge one under each of the $2N_f$ $U(1)$ symmetries and it has charge
$2(N_c-N_f)$ under the R-symmetry.  Therefore, the exact superpotential
is
\eqn\wsunsym{W_{exact}=a\; \left({\Lambda _{N_c,N_f}^{3N_c-N_f}\over
\det_{ff'}(Q_{cf}\tilde Q^{cf'})}\right) ^{1/(N_c-N_f)},}
where $a$ is a constant.  For a single gauge group, our use of the
additional symmetry which is broken by the anomaly (through an
expectation value of $\Lambda$) only gave information which could have
been obtained anyway by using dimensional analysis, as was done in ref.
\dds.  In the examples considered in this paper, however, it will be
crucial for disentangling effects associated with several gauge groups.

The superpotential \wsunsym\ only makes sense for $N_f<N_c$ \ads : for
$N_f=N_c$ the exponent is infinite and for $N_f>N_c$ the determinant is
(classically) zero since the rank of $Q_{cf}\tilde Q^{cf'}$ is then only
$N_c$.  Therefore for $N_f \ge N_c$ the classical vacuum degeneracy is
not removed quantum mechanically.

For $N_f \ge N_c- 1$ the gauge group can be completely broken by the
expectation value of the squarks.  For $N_f<N_c-1$, there is always an
unbroken $SU(N_c-N_f)$ subgroup.  The superpotential \wsunsym\
picks up a $\ZZ _{(N_c-N_f)}$ phase under shifting the theta angle by
$2\pi$.  This phase labels different, though physically equivalent,
vacua of the theory coming from the spontaneous breaking of a discrete
symmetry (by gaugino condensation) in the low energy $SU(N_c-N_f)$
theory.

\subsec{Dynamics: instantons or gaugino condensation}

We now review the dynamics \ads\ leading to the superpotential \wsunsym.
In the case where $N_f=N_c-1$, the gauge group is completely Higgsed
and so instanton methods are reliable.  The $\Lambda $ dependence of
\wsunsym\ indicates that the superpotential for this case is associated
with a single instanton in the completely Higgsed $SU(N_c)$.  An
explicit instanton calculation leads to \wsunsym\ with a non-zero
coefficient $a$ \ads .  It turns out to be natural to define the scale
$\Lambda _{N_c,N_c-1}$ so that $a$=1 in this case.  To relate this
$\Lambda$ to, say, $\Lambda _{\overline{MS}}$ requires a detailed
instanton calculation.  Fortunately, such information is unnecessary for
our purposes.

Having defined our normalization convention for the case of $N_c-1$
flavors, the constant $a$ in \wsunsym\ can be determined for all
$N_f<N_c$ by adding mass terms for $N_c-N_f-1$ of the flavors and
integrating them out.  The symmetries and holomorphy imply that the
exact superpotential for the theory with the mass terms is
\eqn\wsunmass{W_{exact}={\Lambda _{N_c,N_c-1}^{2N_c+1}\over \det
_{ff'}(Q_{cf}\tilde Q^{cf'})}+\sum _{f=N_f+1}^{N_c-1}m_fQ_{cf}\tilde
Q^{cf}.}
For energy scales below the $m_f$, we integrate out the massive flavors
by solving for them using their equations of motion obtained from
\wsunmass\ and find
\eqn\sunnfw{W_{exact}=\epsilon _{(N_c-N_f)}(N_c-N_f)\left({\Lambda
_{N_c,N_f}^{3N_c-N_f}\over \det (Q\tilde Q)}\right)^{1/(N_c-N_f)},}
where $\epsilon _{(N_c-N_f)}$ is a $\ZZ _{(N_c-N_f)}$ phase and where
now $\det (Q_{cf}\tilde Q^{cf'})$ is taken only over the $N_f$ flavors
in the low energy theory.
The scale $\Lambda _{N_c,N_f}$ in eq. \sunnfw\
of the low energy theory is related to
the scale $\Lambda _{N_c, N_c-1}$ of the high energy theory by
\eqn\sunmatching{\Lambda _{N_c,N_f}^{3N_c-N_f}=\Lambda
_{N_c,N_c-1}^{2N_c+1}\prod _{f=N_f+1}^{N_c-1}m_f}
(here we absorb a possible threshold factor
into our definition of $\Lambda _{N_c,N_f}$).
Note that, as in \linres , we can take
${\partial\over \partial m_f}$ of \sunnfw , using \sunmatching ,
to recover
the expectation values of the fields which have been integrated out:
${\partial W_{exact}\over \partial m_f}$=$\langle Q_{f}\tilde
Q^{f}\rangle$.
At this point, we can forget about the massive flavors which have been
integrated out; the superpotential \sunnfw\ is the exact effective low
energy superpotential for $SU(N_c)$ gauge theory with $N_f$ light
flavors.

For $N_f<N_c-1$ the gauge group is not completely broken along the flat
directions and the dynamics leading to \sunnfw\ is associated with
gaugino condensation in the unbroken $SU(N_c-N_f)$ gauge group rather
than with instantons \ads .  The low-energy $SU(N_c-N_f)$ pure
Yang-Mills theory has a scale $\Lambda _{(N_c-N_f),0}$ which is related
to the scale of the high-energy theory by matching the running coupling
constant at the scale, set by the
order parameter $\Delta$, where the theory gets Higgsed
\eqn\sunhiggsmatch{\left({\Lambda _{N_c,N_f}\over E}\right) ^{3N_c-N_f}
=\left({\Lambda_{(N_c-N_f),0}\over E}\right) ^{3(N_c-N_f)}
\qquad\hbox{at}\quad E=(\Delta)^{1 \over 2N_f}.}
As before, we absorb the order one threshold coefficient into the
definition of $\Lambda_{(N_c-N_f),0}$.  The superpotential \sunnfw\ is
thus given by
\eqn\sunwlowscale{W=\epsilon _{(N_c-N_f)}(N_c-N_f)\Lambda
_{(N_c-N_f),0}^3,}
where $\Lam{(N_c-N_f),0}$ is to be thought of as a function of
$\Delta$ and  the high-energy scale $\Lambda _{N_c,N_f}$.
Using \linresg\ in the low energy $SU(N_c-N_f)$ theory
(so $n$=$3(N_c-N_f)$), \sunwlowscale\ gives the gaugino condensate
\eqn\sungci{\langle S_{SU(N_c-N_f)}\rangle =\epsilon _{(N_c-N_f)}\Lambda
_{(N_c-N_f),0}^3.}
Indeed, superpotential \sunnfw\ with $\Lambda _{N_c,N_f}$ held fixed
is exactly equivalent to the low-energy superpotential, obtained
by inserting \sungci\ into the WZ term
\eqn\sunwzterm{W_{WZ}=\int d^2\theta\; [3(N_c-N_f)
-(3N_c-N_f)]\ln\left({\Delta ^{1\over 2N_f}\over M}\right)\;
S_{SU(N_c-N_f)},}
needed in the low energy theory to correct the beta function as in
\sunhiggsmatch , with $\Lambda _{(N_c-N_f),0}$ held fixed.
Note
that by starting with the instanton-induced superpotential for
$N_f=N_c-1$, which is calculated to be non-vanishing, and integrating
out some of the matter fields, we have derived gaugino condensation
\ref\rusglu{M.A. Shifman and A.I. Vainshtein, Nucl. Phys. B296 (1988)
445.}.

The $\ZZ _{N_c-N_f}$ phase in \sungci\ and \sunnfw\ labels the
physically equivalent vacua of $SU(N_c-N_f)$ Yang-Mills associated with
the spontaneous breaking of the $\ZZ _{2(N_c-N_f)}$ chiral symmetry left
unbroken by instantons down to $\ZZ _2$ by the gaugino condensate.
The vacua are physically equivalent because they are related by a
discrete, non-anomalous, R-symmetry.  In particular, the discrete
$\ZZ _{2(N_c-N_f)}$ R-symmetry under which the squarks
are neutral is anomaly free. The terms in
\sunnfw\ are invariant under this symmetry but, because the
superpotential has R-charge 2, the superpotential picks up
a $\ZZ _{N_c-N_f}$ phase under the symmetry.  Therefore, vacua differing
by the phase in \sunnfw\ are physically equivalent.

Finally note that if we add mass terms for all of the flavors and
integrate them out, \linresg , along with the equations of motion and
the matching condition on the scales, gives
\eqn\sungcii{\langle S_{SU(N_c)}\rangle=
\epsilon _{N_c}\Lambda _{N_c,0}^3,}
with a normalization consistent with \sungci .  Using the equations of
motion from \sunnfw\ plus the added tree-level mass terms, we also find
\eqn\sunkona{m_f\langle Q_{cf}\tilde Q^{cf}\rangle=\epsilon_{N_c}
\Lambda _{N_c,0}^3.}
The equality $m_f\langle Q_{cf}\tilde Q^{cf}\rangle$=$\langle S \rangle$,
seen from \sungcii\ and \sunkona , is also a consequence of the Konishi
anomaly; this provides a non-trivial check on our normalization
conventions.

\subsec{Continuous moduli spaces of inequivalent vacua for $N_f\geq N_c$}

We can describe the theories with $N_f\geq N_c$ by starting with the
theory with $N_f=N_c-1$, ``integrating in'' very massive and thus
decoupled matter, and then reducing the mass terms until the extra
matter appears in the low energy theory.  The central feature of the
theories with $N_f\geq N_c$ is that, even at the non-perturbative level,
they have a moduli space of vacua.

For example, when $N_f=N_c$ we see from eq. \sunnfw\ that no invariant
superpotential exists.  Thus there is a continuum of inequivalent vacua
corresponding to different squark expectation values subject to the
D-flatness conditions.  As discussed in ref. \nati , this moduli space
of vacua differs from the classical space of D-flat vacua.  Classically
the singlets $\Delta =\det _{ff'}(Q_{cf}\tilde Q^{cf'})$, $B=\det
Q_{cf}$, and $\tilde B=\det \tilde Q^{cf}$ satisfy the constraint
$\Delta =B\tilde B$.  However, at the quantum level this is modified
(by instantons) to
\eqn\sunconstr{\eqalign{
\Delta -B\tilde B&=\Lambda ^{2N_c}\cr
\pf V &=\Lambda^4\;\;\; {\rm for }\;\; N_c=2,}}
where for $N_c$=2 the constraint is in terms of the $SU(2)$ singlet
fields $V_{fg}=Q_{cf}Q_{c'g}\epsilon ^{cc'}$, which transform as a {\bf
6} under the $SU(4)_F$ flavor symmetry.

For $N_f=N_c+1$, the quantum moduli space of vacua coincides with the
classical space \nati.  The singularity at the origin in this case is
resolved by having extra light fields come down.

\newsec{Illustrative examples based on $SU(2)_1\times SU(2)_2$ gauge
theory}

In this section we illustrate some of our basic points and techniques in
the context of a class of very simple examples based on $SU(2)_1\times
SU(2)_2$ gauge theory.

\subsec{Matter content $Q=(2,2)$ and $L_{\pm}=(1,2)$}

There
are two independent classical D-flat directions, which can be labeled by
$X=Q^2\equiv \half Q_{\alpha \beta}Q_{\gamma \delta}\epsilon ^{\alpha
\gamma}\epsilon ^{\beta \delta}$ and $Y=L_{\alpha  +}L_{\beta -}\epsilon
^{\alpha\beta}$.  At generic values of $X$ and $Y$ the gauge group is
completely broken.  At the classical level, for $X=0$ $SU(2)_1$ is
unbroken and for $Y=0$ there is an unbroken diagonal $SU(2)_D$.

The symmetries $U(1)_{Q}$, $U(1)_{L_{\pm}}$ and $U(1)_R$, with charges
assigned as in \cnclanom\ to the scales $\Lambda _1$ and $\Lambda _2$ of
$SU(2)_1$ and $SU(2)_2$, determine the superpotential to be of the form
\eqn\wxy{W={\Lambda_1^5 \over X} f\left( {\Lambda_2^4 \over XY}\right) .}
Note that for $\Lambda _1\rightarrow 0$ the superpotential goes to zero,
which is the proper behavior for the $SU(2)_2$ gauge theory with four
doublets, as discussed in the previous section.

In order to determine the function $f(u={\Lambda_2^4 \over XY})$ we
first study the limit $u\rightarrow 0$. A term in $f$ proportional to
$u^n$ has a $\Lambda _1$ and $\Lambda _2$ dependence characteristic of
an $SU(2)_1 \times SU(2)_2$ effect with instanton charges $(1,n)$.
Because the gauge group is completely broken, we only expect
contributions associated with instantons -- i.e. only terms proportional
to $u^n$ with $n$ integer.  For small $u$, $f$ thus has the expansion
\eqn\conone{f=\sum_{n=0}^\infty a_n u^n.}
If we set $\Lambda _2$=0, the theory is $SU(2)$ with one flavor (two
doublets) and \sunnfw\ gives $a_0$=1.  The term $a_1u$ in \conone\ has
the quantum numbers of a $(1,1)$-instanton; it can be understood as
follows. For $X\gg Y$ the gauge group is broken to the diagonal subgroup
$SU(2)_D$. An instanton in $SU(2)_D$ then gives,
according to \sunnfw , a superpotential $\Lambda _D^5/Y$. Matching the
running coupling constant of the low energy theory,
$g_D^{-2}=g_1^{-2}+g_2^{-2}$, to the high energy ones at $E=X^{1/2}$,
the scales of the low and high energy theories are related by
$\Lambda _D^5=\Lambda _1^5\Lambda _2^4/X^2$ (there is no finite
threshold correction here in our conventions for the scales)
and thus $a_1$=1 in \conone.

To further determine the function $f$ we temporarily set $\Lambda_1=0$.
Then $SU(2)_2$ couples to four doublets and the model has an $SU(4)$
global symmetry.  The massless modes can be expressed in terms of
$X=Q^2$, $Y=L_+L_-$, and two doublets $A_\pm=QL_\pm$.  Classically, these
six fields are constrained by $XY=A_+A_-$.  Quantum mechanically, this
constraint is modified as in \sunconstr\ to
\eqn\quancons{XY-A_+A_- =\Lambda_2^4.}
Now we weakly gauge $SU(2)_1$.  In this limit of $\Lambda _2\gg \Lambda
_1$, the theory is simply $SU(2)_1$ gauge theory with the two doublets
$A_{\pm}$ and the two singlets $X$ and $Y$, satisfying the constraint
\quancons .  For nonzero $A_+A_-$ the $SU(2)_1$ gauge symmetry is
thus completely broken and the light fields are
only $X$ and $Y$.  There is an unbroken gauge symmetry at $A_+A_-$=0
which, because of $SU(2)_2$ instanton effects in \quancons , is at
$XY=\Lambda _2^4$ rather than the classical value of zero.  Therefore,
the superpotential can only be singular at $u$=1.  In particular, since
the gauge symmetry is broken at $XY$=0 the superpotential cannot be
singular there; the function $f(u)$ in \wxy\ must thus satisfy
$\lim _{u\rightarrow \infty}f(u)\leq O({1\over u})$.

The singularity of the superpotential at $u$=1 is given by
\sunnfw\ for $SU(2)_1$ with the two doublets $A_{\pm}$.
We thus have in the limit
\eqn\nlimit{\Lambda _2^2\gg A_+,A_-\gg \Lambda _1,}
\eqn\wutoi{W={\Lambda _1^5\mu \over A_+A_-}={\Lambda _1^5\mu \over
XY-\Lambda _2^4},}
where $\mu$ is a dimensionful normalization factor, needed because
$A_{\pm}$ are not canonically normalized doublets but, rather,
composites.  Comparing with \wxy , it is seen that $\mu$=$Yg(u)$ for some
function $g$ and thus
\eqn\wwg{W={\Lambda _1^5Yg(u)\over XY-\Lambda _2^4}.}
By holomorphy, the
superpotential must be of the form \wwg\ for any values of the fields
$X$ and $Y$
and scales $\Lambda _1$
and $\Lambda _2$.  Finally, we note that the holomorphic function $g(u)$
can not have any  singularities
in the entire complex $u-$plane (including infinity);
therefore, $g(u)$ must be a constant.
Comparing with the known first term in \conone\ at $u$=0, we find $g(u)$=1.
Therefore, the exact superpotential for this theory is
\eqn\finalw{W= {\Lambda _1^5Y \over XY-\Lam{2}^4}.}
The superpotential \finalw\ exactly sums the multi-instanton expansion
\conone .

We can re-derive the superpotential \finalw\ as the solution of a
differential equation by adding mass terms for the matter fields and
integrating them out.  Adding mass terms to the superpotential \wxy ,
holomorphy and the symmetries determine the exact superpotential to be
\def\mx#1{m_{\hbox{\fiverm #1}}}

\eqn\fmass{W={\Lambda _1^5\over X}\; f\left({\Lambda _2^4\over
XY}\right) +\mx{X} X +\mx{Y} Y}
(note that as in eq. \weefsi , this is linear in the couplings
$\mx{X}$ and $\mx{Y}$).
Below the scales set by the masses, we can integrate out the matter
fields to obtain pure-glue $SU(1)_1\times SU(2)_2$ Yang-Mills theory.
The gaugino condensates in this low-energy theory can be expressed in
terms of the high-energy couplings by taking account of the charges of
these couplings
under the $U(1)_{Q}$, $U(1)_{L_\pm}$ and $U(1)_R$ symmetries and
the fact that the condensates must have charge zero under the $U(1)$
symmetries and charge two under $U(1)_R$.  This gives
\eqn\legcmf{\langle S_1\rangle =\epsilon _1(\mx{X}\Lambda
_1^5)^{1/2}f_1\left( {\mx{Y}\Lambda _2^4\over \Lambda _1^5}\right)
\qquad \langle S_2\rangle=\epsilon _2(\mx{X}\mx{Y}\Lambda
_2^4)^{1/2}f_2\left( {\Lambda _1^5\over \mx{Y}\Lambda _2^5}\right) ,}
where $\epsilon _{1,2}=\pm 1$ and $f_1$ and $f_2$ are functions.
In the limits of large $\mx{X}$ or small $\Lambda
_2$, we can reliably determine $\langle S_1\rangle$ by using
\sungcii\ in the low-energy $SU(2)_1$ Yang-Mills theory and matching the
low-energy scale to our high-energy scales; this gives a condensate as
in \legcmf\ with $f_1$=1.  Since
the argument of $f_1$ is independent of $\mx{X}$, the function $f_1$=1
identically.  Similarly, we can reliably determine that the condensate
$\langle S_2\rangle$ must be independent of $\Lambda_1$ in the limit of
large $\mx{X}$ and  hence $f_2$ must be a constant. The limit where
$\mx{Y}$ is also large determines $f_2$=1.    Thus
\eqn\legcm{\langle S_1\rangle =\epsilon _1(\mx{X}\Lambda
_1^5)^{1/2}
\qquad \hbox{and}\qquad\langle S_2\rangle=\epsilon _2(\mx{X}\mx{Y}\Lambda
_2^4)^{1/2}.}
We can use these
equations together with the (assumed) relations of eq. \wdyndif
\eqn\yvari{\Lambda _1^5{\partial \langle W\rangle \over \partial \Lambda
_1^5}=\langle S_1\rangle \qquad \Lambda _2^4{\partial \langle W\rangle
\over \partial \Lambda _2^4} =\langle S_2\rangle ,}
where, as in sect. 2, $\langle W\rangle$ means the superpotential
\fmass\ with $X$ and $Y$ integrated out -- {\it i.e.} replaced with the
solutions $\langle X\rangle$ and $\langle Y\rangle$ to their equations
of motion, obtained from \fmass\ as functions of the couplings. By varying
$\mx{X}$ and $\mx{Y}$, we can change the expectation values and thereby
determine the function $f$ for all values of its argument.  In
particular, writing the $X$ and $Y$ equations of motion obtained from
\fmass\ as
$$\mx{X}={\Lambda _1^5\over X^2}(f(u)+uf'(u)),
\qquad \mx{Y}={\Lambda _1^5\over \Lambda _2^4}u^2f'(u),$$
a comparison of  \legcm\ and \yvari\ with the superpotential \fmass\
gives differential equations for the function $f(u)$:
$$f^2=(f+uf')\qquad \hbox{and}\qquad f'=(f+uf'),$$
which uniquely determine $f=1/(1-u)$ and thus, in agreement with \finalw ,
$$W={\Lambda _1^5Y\over XY-\Lambda _2^4}.$$
This agreement can be used as further evidence for the assumption \linresg.

We also note that we can take our result \finalw\ and ``integrate in''
the massive fields $S_1$ and $S_2$.  The superpotential which satisfies
\wdyndif\ and which gives \finalw\ upon integrating out $S_1$ and $S_2$
is
\eqn\qlsw{W=S_1\left[\ln \left({\Lambda _1^5\over S_1X}\right)+1\right]+
S_2\ln \left({\Lambda _2^4\over XY}\right)
+S_1\ln \left({S_1+S_2\over S_1}\right)
+S_2\ln \left({S_1+S_2\over S_2}\right).}
The first two terms would be expected following the analysis of \cerne\
for the $SU(2)_1$ and $SU(2)_2$ theories.  The second two terms indicate
the ``interaction'' between the two gauge groups.  A suggestive way to
write \qlsw\ is as
\eqn\glswii{W=S_1\left[\ln \left({\Lambda _1^5\over S_1^2}\right)
+1\right]+S_2\ln \left({\Lambda _2^4\over S_2^2}\right)
+(S_1+S_2)\ln\left({S_1+S_2\over X}\right) +S_2\ln \left({S_2\over
Y}\right) .}
The first two terms in \glswii\ can be associated purely with $SU(2)_1$
and $SU(2)_2$, respectively.  The third term is associated with the
matter field $Q=({\bf 2,2})$ and the fourth is associated with
the $L_{\pm}$.
The expression \glswii\ naturally generalizes, as we will discuss.

\subsec{Matter content $Q=(2,2)$}

If we add a mass term to \finalw\
\eqn\masswex{W=  {\Lambda _1^5Y \over XY-\Lambda_2^4} + \mx{Y}Y ,}
we can integrate out $L_{\pm}$ to obtain the superpotential for an
$SU(2)_1\times SU(2)_2$ theory with matter content $Q=({\bf
2,2})$. $Y$ is easily integrated out; there are two solutions to its
equation of motion leading to
\eqn\weff{W_{eff}= {1 \over X}\left(\Lambda_1^5 \pm 2(\Lambda_1 \tilde
\Lambda_2)^{5/2} + \tilde\Lambda_2^5\right)
= {\left(\Lambda_1^{5/2} \pm\tilde\Lambda_2^{5/2}\right)^2 \over X},}
where the low-energy scale is matched to the high-energy one by $\tilde
\Lambda_2 =(\mx{Y}\Lambda_2^4)^{1/5}$.  So the superpotential for the
$SU(2)_1\times SU(2)_2$ theory with matter content $Q=({\bf 2,2})$ is
\weff ; having integrated out $L_{\pm}$, we can forget about the
original high-energy theory and thus drop the tilde on $\Lambda _2$.

The terms in \weff\ have a clear interpretation.  Along the flat
direction labeled by $X$, the $SU(2)_1\times SU(2)_2$ gauge symmetry
is broken by the Higgs mechanism down to a diagonally embedded
$SU(2)_D$.  An instanton in the broken $SU(2)_1$ gives, according
to \sunnfw , $\Lambda _1^5/X$.  Likewise, an instanton in the broken
$SU(2)_2$ gives $\Lambda _2^5/X$.  Finally, gaugino condensation
in the unbroken $SU(2)_D$ gives the superpotential \sunwlowscale\ (with
the factor of $N_c-N_f$ in \sunwlowscale\
replaced with 2 because the unbroken gauge group is
$SU(2)$) which is
$\pm 2\Lambda _D^3$=$\pm 2\Lambda _1^{5/2}\Lambda _2^{5/2}/X$.  These
are precisely the terms found in our exact answer \weff .

The $\pm$ sign in \weff\ is a discrete label which labels two physically
inequivalent groundstates of the theory.  This sign comes from the fact
that the low energy theory has a $\ZZ _4$ symmetry which is
spontaneously broken down to $\ZZ_2$ by gaugino condensation in the low
energy $SU(2)_D$: $\langle \lambda \lambda \rangle _{SU(2)_D}=\pm
\Lambda _D^3 =\pm \Lambda _1^{5/2}\Lambda _2^{5/2}/X$.  Because of the
contributions of $SU(2)_1$ and $SU(2)_2$ instantons to the
superpotential, the sign choice involved in gaugino condensation in
$SU(2)_D$ label physically inequivalent vacua.  For example, the
potential energy as a function of $X$ differs for the two sign choices
in \weff .  Just as the $SU(N_c)$ theories with $N_f\geq N_c$ have a
continuum of physically inequivalent vacua, this theory has a discrete
choice of physically inequivalent vacua.

To further illuminate these two inequivalent vacua, we add
a mass term for the field $X$ and consider integrating it out.
Using the symmetries, the gaugino condensates in the low-energy
$SU(2)_1\times SU(2)_2$ Yang-Mills theory are of the form
\eqn\fourgrnf{ \langle S_1\rangle = \epsilon _1
\mx{X}^{1/2}\Lam{1}^{5/2}f_1\left({\Lambda _2\over \Lambda _1}\right)
\qquad\hbox{and}\qquad \langle S_2\rangle = \epsilon _2
\mx{X}^{1/2}\Lam{2}^{5/2}f_2\left({\Lambda _1\over \Lambda _2}\right) ,}
where $\epsilon _1$ and $\epsilon _2$ are $\pm 1$ and $f_1$ and $f_2$
are functions.  In the limit of large $\mx{X}$ we can
reliably compute the condensates in \fourgrnf\ by using
\sungcii\ in the low energy Yang-Mills theory and matching the
low-energy scale to the scales of the high-energy theory which includes
the massive field $Q$; this gives $f_1$=1 and $f_2$=1.
Thus, there
are four ground states given by the condensates
\eqn\fourgrn{ \langle S_1\rangle = \epsilon _1
\mx{X}^{1/2}\Lam{1}^{5/2}
\qquad\hbox{and}\qquad \langle S_2\rangle = \epsilon _2
\mx{X}^{1/2}\Lam{2}^{5/2}.}
In the
pure-glue $SU(2)_1\times SU(2)_2$ theory all four states would be
related by a symmetry.  Here, the two states with $\epsilon
_1$=$\epsilon _2$ are indeed related by the spontaneously broken $\ZZ
_4$ symmetry of $SU(2)_D$.  Likewise, the two states with $\epsilon
_1$=$-\epsilon _2$ are related by this symmetry.  On the other hand, the
pair of states with $\epsilon _1$=$\epsilon _2$ are not related by a
symmetry to the pair of states with $\epsilon _1$=$-\epsilon _2$; they
are physically inequivalent.  They differ because of the interactions
with the high energy massive sector.  In particular, the massive field
$Q$ has the expectation value $\mx{X}\langle X\rangle =\epsilon _1
\mx{X}^{1/2}\Lam{1}^{5/2}+\epsilon _2\mx{X}^{1/2}\Lam{2}^{5/2}$.

Another way to
understand this is the following.  In the low energy theory we can perform
independent rotations of the two $\theta$ parameters.  The four ground
states are related by $\theta_i \rightarrow \theta_i + 2\pi$
({\it i.e.} $\tilde
\Lambda_i^6 \rightarrow e^{2\pi i} \tilde\Lambda_i^6$).  In the full
theory which includes the field $Q$, the combination $\theta _1+\theta
_2$ can be rotated away but $\theta _1-\theta _2$ is physical.
Therefore, the two pairs of states related by simultaneous shifts of the
two theta parameters $\theta_i \rightarrow \theta_i + 2\pi$
({\it i.e.} $\Lambda_i^5 \rightarrow e^{2\pi i} \Lambda_i^5$)
are related by
a symmetry but if only one of the $\theta$ parameters is shifted by
$2\pi$ inequivalent ground states are interchanged.

The low energy space includes both the continuous field $Q^2$ and a
discrete label \hbox{$\epsilon_1\epsilon_2=\pm 1$} which determines
the sign in the superpotential.

If we integrate the massive fields $S_1$ and $S_2$ into our
superpotential \weff\ we obtain
\eqn\qsw{\eqalign{W=&S_1\left[\ln\left({\Lam{1}^5\over S_1X}\right)
+1\right]
+S_2\left[\ln\left({\Lam{2}^5\over S_2X}\right)+1\right]\cr
&+S_1\ln \left({S_1+S_2\over S_1}\right)
+S_2\ln \left({S_1+S_2\over S_2}\right).}}
This, again, can be written in the suggestive form
\eqn\qswii{W=S_1\left[\ln\left({\Lambda _1^5\over S_1^2}\right)+1\right]
+S_2\left[\ln\left({\Lambda_2^5\over
S_2^2}\right)+1\right]+(S_1+S_2)\ln \big({S_1+S_2\over X}\big),}
corresponding to terms associated with $SU(2)_1$, $SU(2)_2$, and the
matter field $Q$.

\subsec{Matter content $Q=(2,2)$ and $L_i=(1,2)$ for $i=1\ldots 4$}

The basic gauge singlets are $X=Q^2$ and $V_{ij}=L_iL_j$.  Under the
$SU(4)_F$ flavor symmetry which rotates the $L_i$, $V_{ij}$ transforms
as a {\bf 6}.  Since our superpotential must be built from $SU(4)_F$
singlets, it can only involve $X$ and $\pf V$.  Using the $U(1)_Q$,
$U(1)_{L_i}$ and $U(1)_R$ symmetries, with the scales $\Lambda _1$ and
$\Lambda _2$ assigned charges in accordance with \cnclanom , the exact
superpotential is determined to be of the form
\eqn\wqflf{W={\Lambda _1^5\over X}f\left( u={\Lambda _1^5\Lambda _2^3\over
X^2 \;\pf V}\right).}
In the limit $\Lambda _2\rightarrow 0$ we expect to find a
superpotential corresponding to $f=1$, coming from an instanton in
$SU(2)_1$.  On the other hand, for $\Lambda _1\rightarrow 0$ the theory
is $SU(2)_2$ with six doublets so there is a moduli space of vacua with
a singularity at the origin, corresponding to the fact that there are
extra light fields there \nati .

In order to determine the superpotential, we begin with $\Lambda
_1$=0.  The theory is then $SU(2)_2$ with the six doublets (three
flavors) $Q_{\alpha}$ and $L_i$, where the flavor indices $\alpha =1,2$
and $i=1\dots 4$.  There is a global flavor $SU(6)_F$; the basic
$SU(2)_2$ gauge singlet $U$ transforms as the {\bf 15} of $SU(6)_F$.  In
terms of our original fields, $U$ has the components
$U_{\alpha\beta}=X\epsilon _{\alpha \beta}$, $U_{ij}=V_{ij}$, and
$U_{\alpha i}$.  As  in ref. \nati , all fifteen fields in $U$ are
physical fields in the spectrum.  A superpotential is dynamically
generated which gives six of these fields masses along a flat direction:
\eqn\iidyn{W_{SU(2)_2,dyn}=-{\hbox{Pf}_6U\over \Lambda
_2^3}=-{(X\;\pf V +\Gamma \cdot V)\over \Lambda _2^3},}
where $\hbox{Pf}_6$ is a Pfaffian over the $SU(6)$ indices, $\pf$ is
taken over the $SU(4)$ indices, $\Gamma _{ij}~\equiv~U_{\alpha
i}U_{\beta j}\epsilon ^{\alpha \beta}$, and $\Gamma \cdot V\equiv
{1\over 4}\epsilon ^{ijkl}\Gamma_{ij}V_{kl}$.

We now gauge $SU(2)_1\subset SU(6)_F$, labeled by the index
$\alpha$, keeping $\Lambda _2\gg \Lambda _1$.  Below the scale $\Lambda
_2$, our spectrum consists of the 15 fields $U$ with the superpotential
\iidyn .  The seven composite fields $U_{\alpha\beta}$ and $V_{ij}$ are
$SU(2)_1$ singlets and the fields $U_{\alpha i}$ are four $SU(2)_1$
doublets.  Thus this is the situation \sunconstr\
where there is a moduli space for the scalar components of the
$U_{\alpha i}$ with a constraint which is modified
by a single $SU(2)_1$ instanton to be
\eqn\gammaconstraint{\pf \Gamma =\Lambda _1^5 \Lambda _2^3,}
where we again define $\Gamma _{ij} =\epsilon ^{\alpha \beta}U_{\alpha
i}U_{\beta j}$.
The right hand side of \gammaconstraint\ follows from the symmetries
up to a function of $u$. Inspection of various limits along with
holomorphy implies {\it a posteriori}
 that this function must be unity; for
simplicity then we will not retain it in the following.
The constraint \gammaconstraint\ can be
implemented by a superpotential with a Lagrange multiplier field $A$
\eqn\idyn{W_{SU(2)_1,dyn}=A(\pf \Gamma -\Lambda _1^5\Lambda _2^3).}

Putting together the $SU(2)_1$ and $SU(2)_2$ contributions
\idyn\ and \iidyn\ to the superpotential, we obtain
\eqn\puttogether{W=-{(X\;\pf V+\Gamma \cdot V)\over \Lambda _2^3}
+A(\pf \Gamma -\Lambda _1^5\Lambda _2^3).}
Along the flat direction labeled by an expectation value for $V$, the
superpotential \puttogether\ gives masses to the fields $U_{\alpha i}$
which were not in our original list of fields.  Thus, away from $V$=0, we
can integrate the field $\Gamma$ out of \puttogether .  Upon integrating
out $A$ to implement the constraint on $\pf \Gamma$, the $\Gamma$
equation of motion gives $\langle \Gamma \cdot V\rangle =\pm
2\sqrt{\Lambda _1^5\pf V/\Lambda _2^3}$ and \puttogether\ becomes
\eqn\wintouti{W={-X\;\pf V\over \Lambda _2^3}\pm 2\sqrt{{\Lambda
_1^5\;\pf V\over \Lambda _2^3}}.}
The $\pf V$ equation of motion obtained from
\wintouti\ gives
\eqn\wintoutpfv{W={\Lambda _1^5\over X}\qquad \hbox{and}\qquad \langle
\pf V \rangle ={\Lambda _1^5\Lambda _2^3\over X^2}.}
Result \wintoutpfv\ gives the correct superpotential \sunnfw\ for
$SU(2)_1$ with its one flavor and the correct constraint \sunconstr\ for
$\pf V$ in the limit of large $X$, where the theory is broken to
$SU(2)_D$ with $\Lambda _D^4=\Lambda _1^5\Lambda _2^3/X^2$. Using
holomorphy, eq. \puttogether\ is thus the exact superpotential for this
theory. The complicated looking dynamics in \wintouti\ arises simply
{}from having integrated out the extra fields in $\Gamma$.

The massive fields $S_1$ and $S_2$ can be integrated in, as in the
previous examples.  The result is
\eqn\qlivgsw{W=S_1\left[\ln\left({\Lambda _1^5(X\pf V+\Gamma \cdot
V)\over \pf \Gamma(S_2-S_1)}\right)+1\right]+S_2\left[\ln\left({\Lambda
_2^3(S_2-S_1) \over X\pf V+\Gamma \cdot V}\right)-1\right].}
If we integrate $\Gamma$ out of \qlivgsw\ using the equations of motion
$$\langle X\pf V+\Gamma\cdot V\rangle
=\left({S_2-S_1\over S_2+S_1}\right)X\pf V$$
$$\langle\pf\Gamma\rangle ={S_1^2\over (S_1+S_2)^2}\; X^2\pf V ,$$
eq. \qlivgsw\ becomes
\eqn\qllsw{\eqalign{
W=&S_1\left[ \ln\left( {\Lam{1}^5\over S_1X}\right) +1\right]
+S_2\left[ \ln\left( {\Lam{2}^3S_2\over X\pf V}\right) -1\right]\cr
&\;\;\;\;\; +S_1\ln\left({S_1+S_2\over S_1}\right)
+S_2\ln\left({S_1+S_2\over S_2}\right)\cr
=&S_1\left[\ln\left({\Lambda_1^5\over S_1^2}\right)+1\right]
+S_2\left[\ln\left({\Lambda_2^3 \over S_2^2}\right)-1\right]
+(S_1+S_2)\ln\left({S_1+S_2\over X}\right)\cr
&\;\;\;\;\; +S_2\ln\left({S_2^2\over \pf V}\right). \cr}}

We can re-derive the exact superpotential \wintouti\ by adding mass terms
for the $L_i$ and requiring the result to agree with those of the
previous sections upon integrating out some of the $L_i$.  We thus
consider
\eqn\wqflfm{W={\Lambda _1^5\over X}\; f\left({\Lambda _1^5\Lambda
_2^3\over X^2 \;\pf V}\right)+m_XX+m_VV.}
Suppose we take $V$ and the mass terms $m_V$ to be of the form
\eqn\vmvis{V=\pmatrix{Yi\sigma _2&0\cr 0&Zi\sigma _2}\qquad
m_V=\pmatrix{m_Yi\sigma _2&0\cr 0&m_Zi\sigma _2}}
with $m_Z>m_Y$.  At the scale $m_Z$ we integrate out $Z$.  We should
then obtain the superpotential \masswex\ in the low energy theory with
only two $L_i$.  Rewriting the scales there in terms of our high-energy
scales here using the matching condition at $m_Z$, \masswex\ becomes
\eqn\intoutz{W={\Lambda _1^5\over X}{1\over 1-{m_Z\Lambda _2^3\over
XY}}+m_XX+m_YY.}
Below the scale $m_Y$ we can also integrate out $Y$; the equation of
motion for $Y$ obtained from \intoutz\ is
\eqn\xyis{X\langle Y\rangle
=\mx{Z}\Lambda _2^3\pm \sqrt{{\mx{Z}\Lambda _1^5\Lambda _2^3\over
m_Y}}.}
Having integrated out $V$, this same result must come from the $V$
equations of motion obtained from \wqflfm .  The flavor $SU(4)_F$
covariant way to write \xyis\ and the analogous equation for $\langle
Z\rangle$ is
clearly
\eqn\demveom{X\langle V\rangle
=\left[ \Lambda _2^3\;\pf m_V\pm (\Lambda _1^5\Lambda
_2^3\;\pf m_V)^{1/2}\right] {1\over m_V}.}
On the other hand, the $V$ equation of motion obtained from \wqflfm\ is
\eqn\veom{X\langle V\rangle =\Lambda _1^5uf'(u){1\over m_V}.}
We know that \demveom\ and \veom\ must agree.  Taking the Pfaffian of
\demveom\ and \veom\ gives
$$u^{3/2}f'=\left({\Lambda _2^3\;\pf m_V\over\Lambda_1^5}\right)^{1/2}
\quad\hbox{and}\quad u^{-1/2}=\left({\Lambda_2^3\;\pf m_V\over
\Lambda_1^5}\right)^{1/2}\pm 1.$$
Comparing we get $f'=u^{-2}\pm u^{-3/2}$, which gives $f=-u^{-1}\pm
2u^{-1/2}$, in agreement with our previous result \wintouti .

Note that the $V$ equation of motion in the theory with $m_V\neq 0$
moves $V$ off of the original constraint manifold, eq. \wintoutpfv, to
\eqn\qmsmodified{\langle \pf V\rangle
= {\Lambda _1^5\Lambda _2^3\over X^2}\left[ 1\pm
\sqrt{{\Lambda _2^3 \;\pf m_V \over \Lambda _1^5}}\right] ^2.}
Also, note that if we integrate out $X$ we are left with a low energy
$SU(2)_2$ theory with the four doublets $L_i$.  The equation of motion
obtained from \wintouti\ upon integrating out $X$ gives $\pf
V=m_X\Lambda _2^3=\tilde \Lambda ^4$, where $\tilde \Lambda $ is the
scale of the low energy $SU(2)_2$ theory, in agreement with \sunconstr .

\newsec{Examples with $SO(5)\times SU(2)$ symmetry and
gaugino condensation}

\subsec{Matter content $F=(4,2)$}

The gauge singlet combination is
$X=\half F_{ir}J^{ij}F_{js}\epsilon^{rs}$, with $J$ the $SO(5)$
invariant tensor $i\sigma_2\times \bigone$.  Along the classical flat
direction labeled by $X$, the gauge group is broken down to
$SU(2)'\times SU(2)_D$ where $SU(2)'\subset SO(5)$ and $SU(2)_D$ is
diagonally embedded.  Since the gauge group is not completely
broken, we expect to find non-perturbative effects associated with
gaugino condensation in the unbroken gauge groups rather than with
instantons.

The discussion of sect. 3 generalizes to other gauge theories very
simply: $2N_c$ is replaced in the various formulae with the index of
the adjoint representation of the gauge group and $2N_f$ is replaced
with the sum of the indices of the matter representations.  In
particular, for $SO(5)$ with two ${\bf 4}$'s we replace $N_c$ with 3
and $N_f$ with 1.  Using the $U(1)_F\times U(1)_R$ symmetries with
\cnclanom\ and its $SO(5)$ analog, the superpotential is found to have
the form
\eqn\fwform{W=\left( {\Lambda _5^8\over X}\right) ^{1/2}\;
f\left(u={\Lambda _2^4\over X^2}\right) .}

Along the flat direction with large $X$, the low energy theory is just
the unbroken $SU(2)'\times SU(2)_D$ Yang-Mills theory with the field $X$.
As in \sunwlowscale ,
gaugino condensation in these two Yang-Mills theories gives a
superpotential $W\simeq\pm 2\Lambda _{SU(2)'}^3\pm 2\Lambda _{SU(2)_D}^3$
with the signs of the two condensates independent and with the scales of
the low energy theories related to the high energy ones by the matching
condition, as in \sunhiggsmatch ,
at the scale $X$ where the theory gets Higgsed:
$\Lam{SU(2)'}^3\simeq\Lambda_5^4/X^{1/2}$ and
$\Lam{SU(2)_D}^3\simeq\Lam{5}^4\Lam{2}^2/X^{3/2}$.
In the small $u$ limit, the superpotential is thus given by \fwform\ with
\eqn\fwsmallu{f(u)=\pm 2 \pm 2u^{1/2} +O(u).}
The overall sign of the
superpotential corresponds to two physically equivalent branches of the
square root $(\Lambda _5^8/X)^{1/2}$ in \fwform .
We can thus take the first sign choice in \fwsmallu\ to
be positive.  The relative sign choice between the first two terms is a
discrete label, associated with massive fields, which is needed in the
low energy theory to specify the groundstate. As we will see, it is
related to a spontaneously broken discrete symmetry.

To further examine the function $f$, let us turn off $\Lambda _5$ for
the moment and go to the region of strong $SU(2)$ coupling.  The basic
$SU(2)$ singlet combinations, $V_{ij}=F_{ir}F_{js}\epsilon^{rs}$, form a
${\bf 6}$ of $SU(4)_F$. When $SO(5)$ is gauged, we decompose this
${\bf 6}$ as $V_{ij}=E_{ij}+\half X J_{ij}$, where $X$ is as defined
above, $J$ is the $SO(5)$ singlet mentioned above, and $E$, satisfying
$\Tr JE =0$, transforms as an $SO(5)$ vector.  The constraint \sunconstr\
yields
\eqn\vecnorm{\pf V=\pf E +{1\over 4}X^2=\Lambda_2^4.}
The vector $E$ breaks $SO(5)$ to an $SO(4)\equiv SU(2)_L\times SU(2)_R$
subgroup.  This is to be compared with the $SU(2)'\times SU(2)_D$,
mentioned above, which is unbroken in the weak coupling regions of
field space.  We see from \vecnorm\
that, because of the modified moduli space associated with $SU(2)$
instantons, $SO(5)$ is unbroken at the two points $X=\pm 2\Lambda _2^2$
rather than at the classical value of zero.  The superpotential
\fwform\ can thus only be singular at
$u$=$1/4$.  In particular, the function $f(u)$ must satisfy
$\lim _{u\rightarrow \infty} f(u)\leq O(u^{-1/4})$ to cancel
the singularity in \fwform\ at $X$=0.

In an $SO(5)$ theory with a single canonically normalized vector $\vec
v$, gaugino condensation in the unbroken $SU(2)_L\times
SU(2)_R$ leads to a dynamically generated superpotential
\eqn\wvec{W=2\langle S_L\rangle +2\langle S_R\rangle =\cases{
{2\Lambda _5^4\over \sqrt{\vec v^2}}&for $\langle S_L\rangle=\langle
S_R\rangle$\cr 0&for $\langle S_L\rangle =-\langle S_R\rangle $;}}
the fact that $\langle S_L\rangle$ and $\langle S_R\rangle$ are $\pm
\half \Lambda _5^4/\sqrt{\vec v^2}$ is required by the normalization
conventions of sect. 3  (the factor $\half$ arises
in the matching conditions at $\vec v^2$).  Our vector $E$ differs from
the canonically normalized vector $\vec v$ by some dimensionful,
field-dependent normalization $\mu _v$; in particular,
\eqn\vecvis{\vec v ^2=-\mu _v^{-2} \pf E=\mu _v^{-2}
({1\over 4}X^2-\Lambda _2^4).}

Suppose that near one of the two
points of unbroken $SO(5)$, $X=2\eta \Lambda _2^2$ where $\eta$ can
be either $\pm 1$, $\lamlam{L}$=$\lamlam{R}$.  Using \wvec\ with \vecvis
, the superpotential behaves in this vicinity as
\eqn\wclose{W(X\sim 2\eta \Lambda _2^2)\sim {\Lambda _5^4 \over
\sqrt{X-2\eta \Lambda _2^2}},}
where we have used the fact that $\mu _v\sim \Lambda _2$ in this regime.
There is a unique holomorphic superpotential with the small $u$ and
large $u$ asymptotics mentioned above and the
singularity structure of eq. \wclose :
\eqn\fwexact{W(X,\eta )={2\Lambda_5^4\over\sqrt{X-2\eta\Lambda_2^2}}.}
The superpotential \fwexact\ is thus the exact effective
superpotential for the theory.
The phase $\eta$ appearing in \fwexact\ is a discrete label which,
comparing with \fwsmallu , is the relative sign of the $SU(2)'$ and
$SU(2)_D$ gaugino condensates.

The two choices of groundstates labelled by $\eta$ are
physically equivalent: there is a
discrete $\ZZ_8$ R-symmetry under which $X(\theta)\rightarrow
-X(e^{i\pi/4}\theta )$, which takes $W(X,\eta )
\rightarrow iW(-X,\eta)=W(X, -\eta)$.
For a given value of $\eta$, the superpotential \fwexact\ is singular at
the point \hbox{$X=2\eta\Lambda _2^2$} of unbroken $SO(5)$ but it is
regular at the other point \hbox{$X=-2\eta \Lambda _2^2$} of unbroken
$SO(5)$.  This behavior is possible because of the two branches in
\wvec ; if $\lamlam{L}$=$\lamlam{R}$ near
$X$=$2\eta \Lambda _2^2$, we must have $\lamlam{L}$=$-\lamlam{R}$ near
$X$=$-2\eta \Lambda _2^2$.  The point $X$=$-2\eta\Lambda _2^2$ is
nevertheless singular,
as the normalization $\mu_v$ of the vector $E$ vanishes at this point.

The exact result presented above can be re-derived from \wdyndif\
by adding a mass term for $F$.  With the mass term,
the exact superpotential is determined by the symmetries to be
\eqn\fwformm{W=\left( {\Lambda _5^8\over X}\right) ^{1/2}\;
f\left(u={\Lambda _2^4\over X^2}\right)+\mx{X}X.}
Equations \wdyndif\ give
\eqn\yvarf{\Lambda _5^8{\partial W\over \partial \Lambda
_5^8}=\langle S_5\rangle \qquad \Lambda _2^4{\partial W\over \partial
\Lambda _2^4} =\langle S_2\rangle .}
On the other hand, the gaugino condensates in the low-energy $SO(5)\times
SU(2)$ pure Yang-Mills theories are given by \sungcii\ which, expressed in
terms of the original scales using the matching conditions,
are:\foot{Actually, the second of these equations has been determined
{\it a posteriori}. The symmetries allow the equations in \yvarf\ to be
multiplied by holomorphic functions $f_5$ and $f_2$ of
$\Lambda_5^{8/3}/\mx{X}^{2/3}\Lam{2}^2$.  The function $f_5$ is
determined to be one for large $\mx{X}$ or for small $\Lam{2}$ and is
therefore identically one.  The function $f_2$ is known to be one for
large $\mx{X}$ or small $\Lam{5}$, i.e. only when its argument is small.
However,
the information contained in the first equation
is sufficient for what follows and, indeed, determines that $f_2$=1
identically.}
\eqn\matchfmgc{\langle S_5\rangle =\omega
(\mx{X}\Lam{5}^8)^{1/3}\qquad\hbox{and}\qquad
\langle S_2\rangle =\eta (\mx{X}^2\Lam{2}^4)^{1/2},}
with $\omega^3=1$ and $\eta^2=1$.  Equations \yvarf\ and \matchfmgc\
must agree for every $\mx{X}$. Using the equations of motion from
\fwformm\ to solve for $\mx{X}$, this gives the equations
\eqn\yvarfii{\left({f\over 2}\right)^3=\half f+2uf'\qquad uf'=\eta
\left[u\left(\half f+2uf'\right)^2 \right]^{1/2},}
which uniquely determine
$$f={ 2\over (1- 2\eta u^{1/2})^{1/2}}.$$
So indeed
$$W={2\Lambda _5^4\over (X-2\eta \Lambda _2^2)^{1/2}},$$
as given in eq. \fwexact .

If we integrate the massive fields $S_5$ and $S_2$ into expression
\fwexact\ we obtain the superpotential
\eqn\fssw{\eqalign{
W=&S_5\left[ \ln\left( {\Lam{5}^8\over S_5^2X}\right)+2\right]
+S_2\ln\left( {\Lambda_2^4\over X^2}\right) \cr
&\;\;\;\;\; +S_5\ln\left( {S_5+2S_2\over S_5}\right)
+S_2\ln\left( {(S_5+2S_2)^2\over S_2^2}\right) \cr
= &S_5\left[\ln\left({\Lambda _5^8\over
S_5^3}\right)+2\right] +S_2\ln\left({\Lambda _2^4\over
S_2^2}\right)+(S_5+2S_2) \ln\left({S_5+2S_2\over X}\right) \cr}}
where in the last expression the first two terms look like they
arise from the $SO(5)$ and $SU(2)$ gauge groups and the last term from
the matter field.

\subsec{Matter content $F=(4,2)$ and $L_{\pm}=(1,2)$}

Since in the limit $\Lam{5}\rightarrow 0$ this becomes
$SU(2)$ with six doublets, this
example is similar to that of sect. 4.3.  In particular, we find
interesting behavior near the origin, corresponding to the extra light
fields there.

In terms of the gauge singlet combinations of the superfields
$X$ (as above)
and $Y=L_+^r~L_-^s~\epsilon_{rs}$, the symmetries determine the
exact superpotential to be of the form
\eqn\flwf{W={\Lam{5}^4\over\sqrt{X}}\;
g\left( v={\Lam{5}^4\Lam{2}^3\over X^{5/2} Y}\right) .}
Consider adding a mass term $\mx{Y}Y$ to the superpotential.  In the
limit of large mass, we can integrate out $Y$ to obtain the model of the
previous subsection, a model for which we know the superpotential.
Adding the mass term, we have
$$W={\Lam{5}^4\over\sqrt{X}}\;
g \left( {\Lam{5}^4\Lam{2}^3\over X^{5/2} Y}\right) +\mx{Y}Y.$$
Integrating out $Y$ gives
$$W={\Lambda _5^4\over \sqrt{X}}(g+vg'),$$
where $v$ satisfies $v^2 g'(v)=\mx{Y}\Lambda _2^3/X^2$.  As a function
of $\tilde \Lambda _2^4/X^2$, with $\tilde \Lambda _2^4=m_Y \Lambda
_2^3$ the scale of the $SU(2)$ theory below the scale where $Y$ has been
integrated out, this superpotential must equal that of the $SO(5)\times
SU(2)$ theory with matter field $F=({\bf 4},{\bf 2})$, obtained in sect.
5.1.  Comparing with the result \fwexact , $g$ must therefore satisfy
\eqn\ftyintout{{2\over \sqrt{1-2\eta
v\sqrt{g'}}}=g+vg',}
with $\eta =\pm 1$. This equation, along with some regularity
conditions, can be used to determine $g(v)$.  We are only able to
provide a parametric solution to \ftyintout
\eqn\flparmw{g=\half h(5 -h^2) \qquad
v=\half\alpha (h^{-3}-h^{-5}),}
with $\alpha ^{1/2} =\eta$.  The solution $g(v)$ is, then,
independent of the sign choice $\eta$.

Consider expanding \ftyintout\ or \flparmw\ in the region of small $v$:
$g(v)=\sum _n a_n v^n$. The $n^{th}$ term has the
quantum numbers to be associated with $SO(5)\times SU(2)$ ``instantons''
with charges $((n+1)/2,n)$, where terms with fractional instanton
charges are presumably associated with gaugino condensation.  Using
\ftyintout\ or \flparmw\ we find
\eqn\gsmallv{g=2 +v+3v^2+14v^3+O(v^4).}
The first term in \gsmallv\ is exactly what we expect from gaugino
condensation in the $SU(2)'\subset O(5)$ which remains un-Higgsed along
the flat directions.  The second term can be understood along the flat
direction with $X\gg Y$ where the theory is broken down to an $SU(2)_D$
diagonally embedded in $SO(5)\times SU(2)$.  The $SU(2)_D$ theory has
one light flavor and so an $SU(2)_D$ instanton, which is a $(1,1)$
instanton in the high-energy theory, gives a superpotential as in
\sunnfw\ of $\Lambda_D^5/Y$.  Matching $\Lambda _D$ to the high energy
scales, this gives exactly the term $v$ in \gsmallv .  The higher order
terms in \gsmallv\ are associated with more involved dynamics.

We can also expand \flparmw\ for large $v$ (small $h$):
\eqn\glargev{g(v\rightarrow\infty)={5\over 2}(-2v)^{-1/5}
-(-2v)^{-3/5}+\dots .}

Equations for the superpotential equivalent to \ftyintout\ and \flparmw\
can be re-derived by using \wdyndif\ in the theory with mass terms added
for both $X$ and $Y$
\eqn\flwfm{W={\Lam{5}^4\over\sqrt{X}}\;
g\left( v={\Lam{5}^4\Lam{2}^3\over X^{5/2} Y}\right)+\mx{X}X+\mx{Y}Y.}
We require
\eqn\yvarfl{\Lambda _5^8{\partial W\over \partial \Lambda
_5^8}=\langle S_5\rangle \qquad \Lambda _2^3{\partial W\over \partial
\Lambda _2^3}=\langle S_2\rangle ,}
where the gaugino condensates, expressed in terms of the scales of the
high-energy theory are given by \foot{Again, the second of these
equations is determined {\it a posteriori}.}
\eqn\flgluecond{\langle S_5\rangle =\omega (\mx{X}\Lam{5}^8)^{1/3}
,\qquad\langle S_2\rangle =\epsilon (\mx{X}^2\mx{Y}\Lam{2}^3)^{1/2},}
with $\omega ^3=1$ and $\epsilon ^2$=1.
Using the equations of motion obtained from \flwfm ,
$$\mx{X}={\Lam{5}^4\over 2X^{3/2}}(g+5vg')
,\qquad \mx{Y}={X^2\over\Lam{2}^3 }v^2g',$$
and requiring \yvarfl\ and \flgluecond\ to agree, we obtain equations
which may be written as
\eqn\fleqns{\eqalign{h^3-2vg'=h, \cr
h^6=g'(v),}}
where we define $2h^3=g+5vg'$.  These equations imply the parametric
solution \flparmw .

The massive fields $S_5$ and $S_2$ can be integrated into this theory
yielding
\eqn\flswis{\eqalign{
W=&S_5\left[ \ln\left( {\Lam{5}^8\over S_5^2X}\right) +2\right]
+S_2\left[ \ln\left( {\Lam{2}^3S_2\over X^2Y}\right) -1\right] \cr
&\;\;\;\;\; +S_5\ln\left( {S_5+2S_2\over S_5}\right)
+S_2\ln\left( {(S_5+2S_2)^2\over S_2^2}\right)\cr
=&S_5\left[ \ln\left( {\Lam{5}^8\over S_5^3}\right) +2\right]
+S_2\left[ \ln\left( {\Lambda _2^3\over S_2^2}\right) -1\right]
+(S_5+2S_2)\ln\left( {S_5+2S_2\over X}\right) \cr
&\;\;\;\;\; +S_2\ln \left( {S_2\over Y}\right).}}
Integrating $S_5$ and $S_2$ out of \flswis\ gives
\eqn\flswisev{W=2\langle S_5\rangle -\langle S_2\rangle
\quad\hbox{with}\quad \langle S_{5,2}\rangle ={\Lambda _5^4\over \sqrt
X}h_{5,2}(v),\ h_2 =vh_5^6,\ h_5^2=1+2vh_5^5.}
The superpotential \flswisev\ is seen to be equivalent to \flwf , with
the parametric equation \flparmw\ for $g$ with $h_5$=$h$.

\newsec{Conclusions}

To conclude, some of the non-trivial, non-perturbative dynamics involved
in supersymmetric gauge theories can be explored by a study of their
superpotentials.  Symmetries, holomorphy, and decoupling of heavy fields
provide powerful tools which can often be used to obtain highly
non-trivial superpotentials exactly.

We have demonstrated the power of these techniques in a variety of models.
Some of our techniques, for example adding mass terms to decouple
fields, are particular to theories with matter fields in real
representations of the gauge group.  Others are more general.

We have discussed the unusual procedure of ``integrating in'' --
adding massive
fields to the low energy theory.  Usually, such a procedure is ambiguous
because there are many theories with a massive field leading to the same
low energy theory.  With the assumption that the theory with the massive
field is linear in its source the ambiguity is resolved.  In all of our
examples this assumption was true.

Using this assumption we could also
integrate in the fields $S_s$.  We noticed that in all our examples the
resulting superpotential is of the form
\eqn\conj{W=\sum_sS_s\left[\ln\left({\Lambda _s^{(3G_s-\mu _s)/2}\over
S_s^{G_s/2}}\right)+\half (G_s-\mu _s)\right]
+\sum_t F_t\ln\left({F_t^{d_t}\over Y_t(X^r)}\right),}
where $G_s$ is the index of the adjoint of gauge group $\gG _s$ and
$\mu_s=\sum _i\mu_i^s$, with $\mu_i^s$ the index of the
representation $R_i^s$ of the matter field $\phi _i$
in the gauge group $\gG_s$.
$Y_t(X^r)$ are polynomials in $X^r$ which are invariant under all the
non-Abelian global symmetries ({\it e.g.} $\pf V$ in the example of
sec. 4.3). $F_t$ are linear combinations of the $S_s$ satisfying
$\sum_t q_i(Y_t) F_t = \sum_s \mu_i^s S_s$, where $q_i(Y_t)$ is the
$U(1)_i$ charge of $Y_t$ and $d_t=\half\sum_i q_i(Y_t)$. The first
term in \conj\ can be interpreted as
arising from the gauge sector and the second term is from the matter
fields.  It is easy to check that \conj\ is invariant under all the
global symmetries, including the anomalous ones, and leads to the
Konishi anomaly, eq. \konishi. Clearly, we do not have a proof
of eq. \conj.  However,
given that it was observed to be satisfied in a variety of examples, we
conjecture that it is true under some wide range of
circumstances, thus generalizing the effective Lagrangians of ref. \cerne.

It should be noted that eq. \conj\ is sometimes of limited use.
In some models it is valid but only if more fields $X^r$
than those which are obvious from the classical flat directions
are included. Also, symmetry considerations might not be powerful
enough to determine the polynomials $Y_t$ and $F_t$. In these cases,
eq. \conj\ is still correct but additional dynamical information,
along the lines presented in this paper, is necessary to obtain
the correct superpotential.

Several of the phenomena which we have observed and the tools which we
have used are similar to those which have been encountered in
two-dimensional $N$=2 supersymmetric field theories.
For example, our superpotentials are
sometimes given by an infinite sum over instantons similar to the Yukawa
couplings in Calabi-Yau compactifications.  One of the techniques which
allowed us to perform the sum is the use of differential equations.
These are somewhat reminiscent of the differential equation of ref.
\ref\candelas{P. Candelas, P. Green, L. Parke, and X. de la Ossa,
Nucl. Phys. B359 (1991) 21.}
and the $tt^*$ equations of ref.
\ref\cv{S. Cecotti and C. Vafa, Nucl. Phys. B367 (1991) 359.}.
Also, the fact that we can ``integrate in'' fields is similar to the
situation in 2d gravity coupled to minimal model matter where the KdV
equation allows one to ``flow up'' the renormalization group trajectory.
Since all these two dimensional phenomena are related to an underlying
topological field theory, it is natural to conjecture that our exact
results also have topological interpretations.

Although our techniques rely crucially on supersymmetry, we hope that
the exact results we obtain will be useful in gaining general insight
concerning the dynamics of four dimensional gauge theories.  Finally, it
is worth mentioning that exact results about the superpotentials of
supersymmetric gauge theories are also essential for finding a viable
model of dynamical supersymmetry breaking.

\vskip36pt
\centerline{{\bf Acknowledgements}}

We would like to thank T. Banks, M. Dine, M. Douglas, G. Moore,
P. Pouliot, S. Shenker, and E. Witten for useful discussions.
This work was supported in part by DOE grant \#DE-FG05-90ER40559.

\listrefs

\end